\documentclass{emulateapj}
\usepackage{xcolor}

\received{January 1, 0000}
\revised{January 1, 0000}
\accepted{\today}
\shorttitle{New Luminosity--Temperature Relation}
\shortauthors{Fujita \& Aung}

\begin{document}

\title{Halo Concentrations and the New Baseline X-ray Luminosity-Temperature and Mass Relations of Galaxy Clusters}

\author{Yutaka Fujita\altaffilmark{1}, Han Aung\altaffilmark{2}}

\affil{
$^1${Department of Earth
and Space Science, Graduate School of Science, Osaka University,
Toyonaka, Osaka 560-0043, Japan}\\
$^2${Department of Physics, Yale University, New Haven, CT 06520, USA}
}
\email{Email: fujita@vega.ess.sci.osaka-u.ac.jp}

\begin{abstract}
The standard self-similar model of galaxy cluster formation predicts that the X-ray
luminosity--temperature ($L_{\rm X}$--$T_{\rm X}$) relation of galaxy clusters
should have been $L_{\rm X}\propto T_{\rm X}^2$ in absence of the baryonic physics,
such as radiative cooling and feedback from stars and black holes.
However, this baseline relation is predicted without considering the
fact that the halo concentration and the characteristic density of
clusters increases as their mass decreases, which is a consequence of
hierarchical structure formation of the universe. Here, we show that the
actual baseline relation should be $L_{\rm X}\propto T_{\rm X}^\alpha$, where
$\alpha\sim 1.7$, instead of $\alpha=2$, given the mass dependence of
the concentration and the fundamental plane relation of galaxy
clusters. Numerical simulations show that $\alpha\sim 1.6$, which is
consistent with the prediction. We also show that the baseline
luminosity--mass ($L_{\rm X}$--$M_\Delta$) relation should have been
$L_{\rm X}\propto M_\Delta^\beta$, where $\beta\sim 1.1$--1.2, in contrast
with the conventional prediction ($\beta=4/3$). In addition, some of the scatter in the $L_{\rm X}$--$M_\Delta$ relation can be attributed to the scatter in the concentration--mass ($c$--$M$) relation. The confirmation
of the shallow slope could be a proof of hierarchical clustering. As an
example, we show that the new baseline relations could be checked by
studying the temperature or mass dependence of gas mass fraction of
clusters. Moreover, the highest-temperature clusters would follow the
shallow baseline relations if the influences of cool cores and cluster
mergers are properly removed.
\end{abstract}


\section{Introduction} 
\label{sec:intro}

Clusters of galaxies have grown from slightly overdense regions of
the universe. The initial density fluctuations of the universe are
described by a random Gaussian field with a power spectrum having a
smoothly changing power-law index. Thus, clusters are expected to have a
high degree of self-similarity in scale and time, which leads to various
scaling relations among observables, although the relations may be
affected by the mass assembly histories of dark matter halos. The
relation between the X-ray luminosity $L_{\rm X}$ and temperature $T_{\rm X}$ of
clusters has been studied for many years, probably because it is
relatively easy to measure them. Observations have shown that the
relation is approximately described as $L_{\rm X}\propto T_{\rm X}^3$
\citep[e.g.][]{1991MNRAS.252..414E,1998ApJ...504...27M}. This relation
is thought to be influenced by feedback from active galactic nuclei
(AGNs) and supernovae in galaxies in clusters
\citep[e.g.][]{2002ApJ...576..601V,2004MNRAS.348.1078B,2008ApJ...687L..53P}.
The relation when there was no feedback is often expected to be
$L_{\rm X}\propto T_{\rm X}^2$ \citep{1986MNRAS.222..323K,1998ApJ...495...80B}. We
refer to scaling relations when there were no nongravitational effects
(e.g. feedback and radiative cooling) as the ``baseline relations,''
because they are used as baselines when the nongravitational effects
are estimated.

The conventional baseline relation of $L_{\rm X}\propto T_{\rm X}^2$ is derived
assuming that cluster structure is self-similar and the characteristic
density of clusters is proportional to the density of the background
universe. This means that clusters at a given redshift have a common
characteristic density. However, this assumption seems to be at odds
with recent studies on structure formation of the universe. Numerical
simulations have shown that the dark matter density profile of galaxy
clusters is well represented by the Navarro--Frenk--White (NFW) density
profile \citep{1997ApJ...490..493N}, 
and that less massive clusters tend to be more concentrated and
have higher characteristic densities. In other words, the characteristic
density differs among clusters even at a given redshift. This is because
in the standard CDM cosmology, less massive clusters form earlier and
their characteristic density reflects the higher background density of
the universe at their formation time
\citep[e.g.][]{1997ApJ...490..493N,2002ApJ...568...52W,2008A&A...482..451Z,2013MNRAS.432.1103L}.
Thus, their mass dependence is a consequence of hierarchical structure
formation of the universe
\citep[e.g.][]{2008MNRAS.390L..64D,2013ApJ...766...32B,2014ApJ...797...34M}.

Scaling relations for clusters are not necessarily limited to one-to-one
correlations. For example, relations among three parameters are often
considered and ``fundamental planes'' are the representative ones
\citep{1993MNRAS.263L..21S,1998A&A...331..493A,1999ApJ...519L..51F,2002ApJ...581....5V,2004ApJ...600..640L,2006ApJ...640..673O,2009MNRAS.400.1317A,2013MNRAS.435.1265E,2014MNRAS.437.1171M,2015MNRAS.446.2629E}.
Recently, \citet[][see \citealt{2019arXiv190100008F} for a
review]{2018ApJ...857..118F} found that observed clusters are
distributed on a plane in the space of $(\log r_s, \log M_s, \log T_{\rm X})$,
where $r_s$ and $M_s$ are the characteristic radius and mass for the NFW
profile, respectively, and $T_{\rm X}$ is the X-ray temperature. Numerical
simulations have confirmed the plane and have shown that clusters evolve
along the plane. Thus, this fundamental plane reflects the structure and
evolution of dark matter halos of clusters. The nongravitational
effects and cluster mergers have little effect on the plane. The
properties of the plane can be explained by an analytical model of
structure formation constructed by \citet{1985ApJS...58...39B}. In
particular, the angle of the plane in the space of $(\log r_s, \log M_s,
\log T_{\rm X})$ indicates that clusters have not perfectly achieved virial
equilibrium because of continuous matter accretion from the surroundings
\citep{2018ApJ...857..118F}.  We note that the deviation from the virial
equilibrium is not considered when the conventional relation $L_{\rm X}\propto
T_{\rm X}^2$ is constructed.

In this study, we revise the baseline $L_{\rm X}$--$T_{\rm X}$ relation considering
the mass dependence of the halo concentrations and the fundamental
plane.  We also study the baseline luminosity--mass ($L_{\rm X}$--$M_\Delta$)
relation as a corollary. The paper is organized as follows. In
Section~\ref{sec:convential}, we review the derivation of the
conventional baseline relations. In Section~\ref{sec:base}, we derive
the revised baseline $L_{\rm X}$--$T_{\rm X}$ and $L_{\rm X}$--$M_\Delta$ relations by
taking into account the mass dependence of halo concentrations and the
fundamental plane relation, and show that they deviate from the
conventional relations. In Section~\ref{sec:mum}, we test the
predictions of our new model using the \textit{Omega500} hydrodynamical
cosmological simulations of galaxy cluster formation. In
Section~\ref{sec:dis}, we discuss future observations of the
$L_{\rm X}$--$T_{\rm X}$ and $L_{\rm X}$--$M_\Delta$ relations. In Section~\ref{sec:sum},
we summarize our main results.

In this paper, we assume a spatially flat $\Lambda$CDM cosmology with
$\Omega_\mathrm{m}=0.27$, $\Omega_\Lambda=0.73$, and the Hubble constant
of $H_0=100\:h$\,km\,s$^{-1}$\,Mpc$^{-1}$ for $h=0.7$, unless otherwise mentioned.

\section{conventional baseline relations} 
\label{sec:convential}

The conventional scaling relations are based on a gravitational
collapse model of a homogeneous spherical overdense region in the
Einstein-de Sitter universe \citep{1986MNRAS.222..323K}. This region
initially expands with the Hubble expansion. Then, owing to the gravity,
it deviates from the expansion, and starts to collapse. The evolution is
self-similar and can be treated analytically
\citep[e.g.][]{1980lssu.book.....P}. If the collapsed region is
virialized, the average density is $18\: \pi^2\sim 200$ times the
critical density of the universe. Subsequent matter accretion from the
surroundings is not considered in this model.

The conventional baseline relation of $L_{\rm X}\propto T_{\rm X}^2$ can be
derived as follows.  First, we assume that the typical density of
clusters is $\rho_\Delta\equiv \Delta\rho_c(z)$, where $\Delta$ is a
constant and $\rho_c(z)$ is the critical density of the universe at
redshift $z$. The critical density depends on $z$ as in
$\rho_c(z)\propto E(z)^2$, where the Hubble parameter at $z$ is
represented by $H(z)=H_0E(z)$ and $H_0$ is the Hubble constant. The
density $\rho_\Delta$ does not depend on clusters and is constant at a
given redshift. The corresponding cluster radius $r_\Delta$ is defined
as the one inside which the average density is $\rho_\Delta$, and the
mass is written as
\begin{equation}
\label{eq:MD}
 M_\Delta = \frac{4\pi}{3}\rho_\Delta r_\Delta^3\:.
\end{equation}
For the overdensity, $\Delta=200$ is often used because it is close to
$18\: \pi^2$. However, it is generally difficult to observe cluster
properties out to $r_{200}$; $\Delta=500$ is also often used.

Assuming that the cooling function of the intracluster medium (ICM) is
described by bremsstrahlung, the bolometric emissivity is proportional
to $\rho_{\rm ICM}^2 T_{\rm X}^{1/2}$, where $\rho_{\rm ICM}$ is the typical
density of the ICM. Here, we assume that $\rho_{\rm ICM}\propto
\rho_\Delta$. Since the typical volume of a cluster is proportional to
$r_\Delta^3$, the X-ray luminosity of clusters is represented by
\begin{equation}
 \label{eq:Lx_D}
L_{\rm X}\propto \rho_\Delta^2 T_{\rm X}^{1/2}r_\Delta^3\:.
\end{equation}
If we assume the virial equilibrium, the X-ray temperature is given by
$T_{\rm X}\propto M_\Delta/r_\Delta \propto \rho_\Delta r_\Delta^2$ using
equation~(\ref{eq:MD}). Considering that $\rho_\Delta\propto E(z)^2$ and
$r_\Delta\propto T_{\rm X}^{1/2}E(z)^{-1}$, we finally obtain the relation of
\begin{equation}
\label{eq:LTE}
 L_{\rm X}\propto T_{\rm X}^2 E(z)
\end{equation}
from equation~(\ref{eq:Lx_D})
\citep{1986MNRAS.222..323K,1998ApJ...495...80B}. Thus,
$L_{\rm X}\propto T_{\rm X}^2$ for a given $z$. Similarly, the baseline
luminosity--mass relation can be obtained as in $L_{\rm X}\propto
M_\Delta^{4/3}E(z)^{7/3}$ \citep{1998ApJ...495...80B}.

\section{New baseline relations} 
\label{sec:base}

However, the above derivations do not take into account the mass profile
of clusters. Numerical simulations have shown that the dark matter
density profile of galaxy clusters is well represented by the NFW
density profile \citep{1997ApJ...490..493N}:
\begin{equation}
\label{eq:NFW}
 \rho_{\rm DM}(r) = \frac{\delta_c\rho_c}{(r/r_s)(1+r/r_s)^2}\:,
\end{equation}
where $r$ is the cluster centric radius, $r_s$ is the characteristic
radius, and $\delta_c$ is the normalization. The radius $r_s$ is smaller
than $r_\Delta$ for clusters if $\Delta=200$ and~500. We define the
characteristic mass $M_s$ as the mass enclosed within $r=r_s$, and define the characteristic density as $\rho_s\equiv 3\: M_s/(4\pi r_s^3)$. 
The halo concentration parameter is given by
\begin{equation}
 \label{eq:cD}
c_\Delta=r_\Delta/r_s\:.
\end{equation}
The mass profile of the NFW profile is then given by
\begin{equation}
\label{eq:MNFW}
 M(r) = 4\pi\delta_c\rho_c r_s^3
\left[\ln\left(1+\frac{r}{r_s}\right)-\frac{r}{r+r_s}\right]\:.
\end{equation}
From this equation, the characteristic mass $M_s$ can be expressed in terms of $M_\Delta$ and $c_\Delta$:
\begin{equation}
\label{eq:MDMs}
 M_s = M_\Delta\frac{\ln 2-1/2}{\ln(1+c_\Delta)-c_\Delta/(1+c_\Delta)}\:.
\end{equation}

$N$-body simulations have shown that $c_\Delta=c_\Delta(M_\Delta, z)$ is
a decreasing function of $M_\Delta$ for a given $z$, with a considerable
dispersion ($\sim 0.1$~dex) due to the diversity in cluster ages for a
given mass
\citep[e.g.][]{2008MNRAS.390L..64D,2013ApJ...766...32B,2014ApJ...797...34M,2018ApJ...863...37F}.
While a wide mass range of halos have concentration of $c_{200}\sim 3$
at high redshifts, only most massive halos have $c_{200}\sim 3$ and
others have higher concentration at $z\sim 0$
\citep{2018ApJ...859...55C}. As a result, from equations~(\ref{eq:MD})
and~(\ref{eq:cD}), the scale radius $r_s\propto
(M_\Delta/\rho_\Delta(z))^{1/3}/c_\Delta(M_\Delta, z)$ and the
characteristic density $\rho_s= 3\: M_s/(4\pi r_s^3)$ also depend on
$M_\Delta$ and $z$.

Since the emissivity of the ICM is proportional to the density squared,
the X-ray luminosity of clusters should reflect the structure of their
central region where the density is high. If we assume that the
dark matter profile follows the NFW profile (equation~(\ref{eq:NFW}))
and that the ICM density follows that of dark matter ($\rho_{\rm
ICM}\propto \rho_s$), the characteristic volume of a cluster should be
$\propto r_s^3$ and the X-ray luminosity is
\begin{equation}
 \label{eq:Lx_s}
L_{\rm X} \propto \rho_s^2 T_{\rm X}^{1/2}r_s^3\:,
\end{equation}
in contrast with equation~(\ref{eq:Lx_D}). Since $\rho_s$ depends on
$M_\Delta$ and $z$ while $\rho_\Delta$ is constant for a given $z$, this
fact differentiates equation~(\ref{eq:Lx_s}) from
equation~(\ref{eq:Lx_D}). In other words, the variation of the
halo concentration among clusters is not considered in the derivation of
equation~(\ref{eq:LTE}). As clusters with larger $M_\Delta$ tend to
have larger $T_{\rm X}$, we expect that the mass dependence of $\rho_s$
affects the $L_{\rm X}$--$T_{\rm X}$ relation as well as the $L_{\rm X}$--$M_\Delta$
relation if they are derived from equation~(\ref{eq:Lx_s}) (see also \citealp{2001ApJ...556...77E}).

The revised baseline $L_{\rm X}$--$T_{\rm X}$ and $L_{\rm X}$--$M_\Delta$ relations of
clusters can be obtained using the mass dependence of the concentration
parameter $c_\Delta$ and the fundamental plane relation given by
\begin{equation}
\label{eq:TX}
 T_{\rm X} = T_{X0}\left(\frac{r_s}{r_{s0}}\right)^{-2}
\left(\frac{M_s}{M_{s0}}\right)^{(n+11)/6}\:,
\end{equation}
where $(r_{s0}, M_{s0}, T_{X0})$ is a representative point on the
fundamental plane
\citep{2018ApJ...857..118F,2018ApJ...863...37F}.\footnote{We use
$r_{s0}=414$~kpc, $M_{s0}=1.4\times 10^{14}\: M_\odot$, and
$T_{X0}=3.7$~keV based on the results of the MUSIC simulations
\citep{2014ApJ...797...34M,2018ApJ...857..118F}.}  Note that $T_{\rm X}$ is the
core excised temperature. The relation does not depend on $z$ at least
$z\lesssim 1$ and indicates that clusters in general have not achieved
virial equilibrium \citep{2018ApJ...857..118F}. 
The equation~(\ref{eq:TX}) can be derived from the entropy
constant given in the similarity solution by
\citet{1985ApJS...58...39B}; the constant reflects the conservation of
the ICM entropy. The relation depends on the spectral index $n$ of the
density perturbations of the universe, because the overdense region that
later becomes the inner part of a cluster and gives the inner boundary of
the solution of \citet{1985ApJS...58...39B} evolves from the density
perturbations \citep{2018ApJ...857..118F}. Although the index should be
$n\sim -2$ at cluster scales
\citep[e.g.][]{1998ApJ...496..605E,2015ApJ...799..108D}, we treat $n$
simply as a parameter here.  We assume $n=-2$ and~$-2.5$, both of which
are consistent with observed and simulated fundamental planes
\citep{2018ApJ...863...37F}.  Since equation~(\ref{eq:TX}) shows that
$T_{\rm X}$ is a function of $r_s$ and $M_s$, it is also a function of
$M_\Delta$ and $z$.

Although we considered only bremsstrahlung for the cooling function
($\Lambda\propto T_{\rm X}^{1/2}$) in Section~\ref{sec:convential} and in
equation~(\ref{eq:Lx_s}) for simplicity and illustration, we will now
include the effect of metal-line cooling, which introduces additional
dependence on the ICM metallicity $Z$.
The X-ray luminosity would then be given by $L_{\rm X}\propto n_e^2 \Lambda(T_{\rm X},Z)
V_{\rm rs}\propto \rho_s^2 \Lambda(T_{\rm X},Z) V_{\rm rs}$, where $V_{\rm rs}=(4\pi/3)r_s^3$ and we adopt the following metallicity-dependent cooling function $\Lambda$ given by
\begin{eqnarray}
\Lambda(T_{\rm X},Z)&=&2.41\times 10^{-27}
\left[0.8+0.1\left(\frac{Z}{Z_\odot}\right)\right]
\left(\frac{T_{\rm X}}{\rm K}\right)^{0.5}\nonumber\\
&+ & 1.39\times 10^{-16}
\left[0.02+0.1\left(\frac{Z}{Z_\odot}\right)^{0.8}\right]\nonumber\\
& & \times\left(\frac{T_{\rm X}}{\rm K}\right)^{-1.0}\rm\: erg\: cm^{3}\:
\end{eqnarray}
\citep{2013MNRAS.428..599F}, which approximates the cooling function derived by \citet{1993ApJS...88..253S} for $T_{\rm X}\ga 10^5$~K and $Z\la 1\:
Z_\odot$. 

\begin{figure}
\plotone{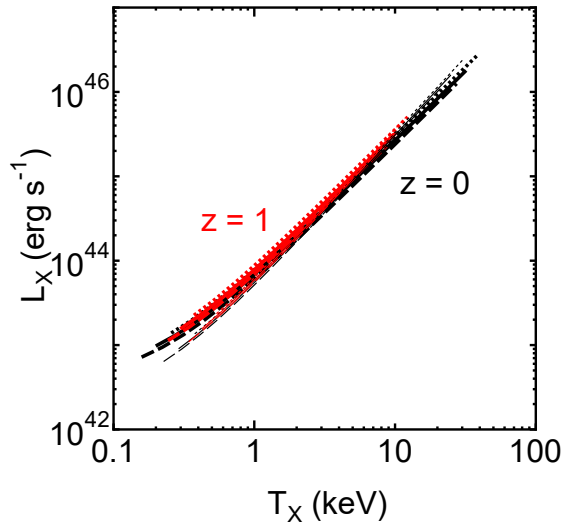} \caption{$L_{\rm X}$--$T_{\rm X}$ relations for $n=-2$ (thick
lines) and $n=-2.5$ (thin lines). Black lines and red lines represent
$z=0$ and~1, respectively.  Solid lines are calculated for the fiducial
$c_\Delta$--$M_\Delta$ relation. The dotted and dashed lines correspond
to $c_\Delta^U$ and $c_\Delta^L$, respectively. The vertical axis
($L_{\rm X}$) is not corrected by $E(z)$.\label{fig:LT}}
\end{figure}

\begin{figure}
\plotone{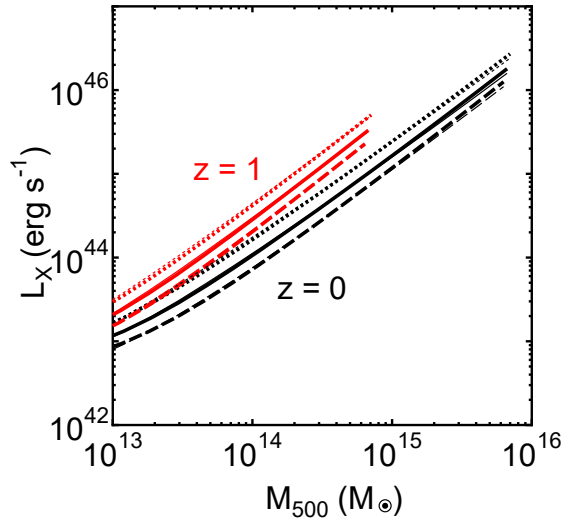} \caption{Same as Figure~\ref{fig:LT}, but for
$L_{\rm X}$--$M_\Delta$ relations for $\Delta=500$. Thick and thin lines are
almost identical.\label{fig:LM}}
\end{figure}

%\begin{deluxetable}{CCCCCC}[b!]
%\tablecaption{The predictions of the indices $\alpha$ and $\beta$\label{tab:index}}
%\tablecolumns{4}
%\tablenum{1}
%\tablewidth{0pt}
%\tablehead{
%\colhead{$n$} &
%\colhead{$\alpha$} &
%\colhead{$\beta_{200}$\tablenotemark{a}} &
%\colhead{$\beta_{500}$\tablenotemark{b}} 
%}
%\startdata
%-2   & 1.59-1.61 & 1.17-1.24 & 1.19-1.26 \\
%-2.5 & 1.74-1.77 & 1.12-1.20 & 1.14-1.22 \\
%\enddata
%\tablenotetext{a}{$\beta$ for $\Delta=200$}
%\tablenotetext{b}{$\beta$ for $\Delta=500$.}
%\end{deluxetable}

\begin{table}[b]
\tabcaption{The predictions of the indices $\alpha$ and $\beta$\label{tab:index}}
\begin{center}
\begin{tabular}{cccc}
\tableline\tableline\noalign{\smallskip}
 $n$ & $\alpha$ & $\beta_{200}^a$ & $\beta_{500}^b$ \\
    &  &  &   \\
\noalign{\smallskip}\tableline\noalign{\smallskip}
-2   & 1.59--1.61 & 1.17--1.24 & 1.19--1.26 \\
-2.5 & 1.74--1.77 & 1.12--1.20 & 1.14--1.22 \\
\noalign{\smallskip}\tableline\noalign{\smallskip}
\end{tabular}
\\
Notes --- $^a$$\beta$ for $\Delta=200$, and $^b$$\beta$ for
$\Delta=500$.
\end{center}
\end{table}

Now, using $M_\Delta$ as a parameter, we can draw the $L_{\rm X}$--$T_{\rm X}$
relation (Figure~\ref{fig:LT}). We assume that $L_{\rm X}=n_e^2 \Lambda(T_{\rm X},
0.3\: Z_\odot) V_{\rm rs}$. Assuming that the ICM consists of hydrogen
and helium\footnote{The inclusion of metals have little
influence (only a few percent change in $n_e$) as long as $Z\lesssim 1\:
Z_\odot$.}, the electron density is given by $n_e = 0.86\:f_{\rm
gas}\rho_s/m_{\rm p}$, where $m_{\rm p}$ is the proton mass and $f_{\rm gas}=0.13$
is the gas mass fraction of massive clusters
\citep[e.g.][]{2006A&A...452...75B,2013ApJ...778...14G,2015MNRAS.450..896D}.
The solid lines are the fiducial relations that are calculated using the
analytic function\footnote{https://bitbucket.org/astroduff/commah} of
$c_\Delta=c_\Delta(M_\Delta, z)$ obtained by
\citet{2015MNRAS.452.1217C}.

Considering the dispersion of the $c_\Delta$--$M_\Delta$ relation, we
also represent the $L_{\rm X}$--$T_{\rm X}$ relations when $c_\Delta$ (fiducial) is
replaced by $c_\Delta^U = 10^{0.1} c_\Delta$ (dotted line) or
$c_\Delta^L = 10^{-0.1} c_\Delta$ (dashed line). Figure~\ref{fig:LT}
shows that the solid, dotted, and dashed lines are almost identical,
which means that the dispersion of the $c_\Delta$--$M_\Delta$ relation
does not introduce a scatter in the $L_{\rm X}$--$T_{\rm X}$ relation. Moreover, the
$L_{\rm X}$--$T_{\rm X}$ relation does not depend sensitively on redshift, which is
in contrast to the relation derived based on the simple self-similar
model on a scale of $r_{\Delta}$ (equation~(\ref{eq:LTE})). Since the
baseline relation does not evolve (Figure~\ref{fig:LT}), observed
evolution, if any, can be attributed to additional baryonic physics,
such as gas cooling and feedback.

Observationally, there seems to be no consensus about the redshift
evolution of the $L_{\rm X}$--$T_{\rm X}$ relation so far
\citep[e.g.][]{2012A&A...539A.120B}, although
\citet{2011A&A...535A...4R} concluded that the evolution of X-ray
luminosity for a given temperature is slower than predicted by a
simple self-similar model (see equation~\ref{eq:LTE}). Note that the
$L_{\rm X}$--$T_{\rm X}$ relation in Figure~\ref{fig:LT} is constructed from
$\rho_s$ and $T_{\rm X}$, which reflect cluster properties on a scale of $r_s$.
The observed $L_{\rm X}$--$T_{\rm X}$ relation has a larger scatter ($\sim 0.2$~dex;
e.g. \citealt{2012MNRAS.421.1583M}) than those shown by the dotted and
dashed lines in Figure~\ref{fig:LT}, which suggests that actual X-ray
luminosities are impacted by local and/or temporary phenomena around the
cluster centers (e.g., AGN feedback in cool cores and/or disruption of
the cores by cluster mergers) and that the effects differ among
clusters.

Table~\ref{tab:index} summarizes the values of the index
$\alpha$ of the $L_{\rm X}$--$T_{\rm X}$ relation ($L_{\rm X}\propto T_{\rm X}^\alpha$) for
different $n$ and a temperature range of $1<T_{\rm X}<10$~keV. For each index,
the smaller one is for $z=0$ and the larger one is for $z=1$. The index
is determined for the fiducial relation (solid lines in
Figure~\ref{fig:LT}) but the results are almost unchanged even if we
take the dotted or dashed lines. Although the slope of the $L_{\rm X}$--$T_{\rm X}$
relation becomes slightly shallower for $T_{\rm X}\lesssim 3$~keV due to the
metal-line cooling (Figure~\ref{fig:LT}), the magnitude of this effect
is quite small. The indices for a temperature range of $3<T_{\rm X}<10$~keV is
larger than those for $1<T_{\rm X}<10$~keV only by $\sim 0.03$. We,
therefore, conclude that the index $\alpha$ should be smaller than two
and should be $\alpha\sim 1.6$--1.8 if the ICM density profile follows
the dark matter profile.  The shallower slope is ascribed to the
increase in the halo concentration $c_\Delta$ and the characteristic
density $\rho_s$ for lower temperature (less massive) clusters. We
expect that the smaller index $\alpha$ is realized when additional
physics such as feedback, radiative cooling, and disturbance by cluster
mergers are ignorable, because the ICM settled in the potential well of
the dark matter halo and $r_s$ is the only spatial scale of the NFW
profile.

It may be instructive to represent the $L_{\rm X}$--$T_{\rm X}$ relation
only by the fundamental plane. From equations~(\ref{eq:Lx_s}),
(\ref{eq:TX}), and $\rho_s\propto M_s/r_s^3$, we obtain $L_{\rm X}\propto
T_{\rm X}^{(19+n)/(14+2n)}\rho_s^{(3+n)/(7+n)}$, which is $L_{\rm X}\propto
T_{\rm X}^{1.7}\rho_s^{0.2}$ for $n=-2$, and $L_{\rm X}\propto
T_{\rm X}^{1.83}\rho_s^{0.11}$ for $n=-2.5$. This means that the $L_{\rm X}$--$T_{\rm X}$
relation is not sensitive to $\rho_s$ and that the $L_{\rm X}$--$T_{\rm X}$ relation
is close to an ``edge-on view'' of the fundamental plane (see also
\citealt{1999ApJ...519L..51F}). Thus, the relation is almost independent
of $z$ and has almost no dispersion (Figure~\ref{fig:LT}(a)), even
though $\rho_s$ is a function of $z$ and $M_\Delta$.\footnote{The
fundamental plane reflects the higher concentration of lower-mass and/or
higher-redshift clusters. The function $c_\Delta=c_\Delta(M_\Delta,z)$
approximately defines the evolution track of clusters on the fundamental
plane (Figure 2 in \citealt{2018ApJ...863...37F}).}

As $L_{\rm X}$ is a function of $M_\Delta$ and $z$, the $L_{\rm X}$--$M_\Delta$
relation can immediately be obtained. Figure~\ref{fig:LM} shows the
$L_{\rm X}$--$M_\Delta$ relation for $\Delta=500$. Table~\ref{tab:index}
reports the index of the relation $\beta$ ($L_{\rm X}\propto M_\Delta^\beta$)
for a temperature range of $1<T_{\rm X}<10$~keV, where $\beta_{200}$ and
$\beta_{500}$ are $\beta$ for $\Delta=200$ and 500, respectively. For
each index, the smallest one is for $z=0$ and the largest one is for
$z=1$. The indices are determined for the fiducial relations, indicated
with the solid lines in Figure~\ref{fig:LM}. The indices for a
temperature range of $3<T_{\rm X}<10$~keV is larger than those for
$1<T_{\rm X}<10$~keV only by $\sim 0.03$. The table shows that $\beta\sim
1.2$, which is slightly smaller than the prediction of the conventional
self-similar model ($\beta=4/3\approx 1.33$;
Section~\ref{sec:convential}). Clusters with $M_{500}\lesssim
10^{14}\: M_\odot$ host a plasma with an X-ray emission with a relatively
larger contribution from metal-line recombination. For a given mass,
$L_{\rm X}$ at $z=1$ is larger than that at $z=0$, mainly because the
characteristic density $\rho_s$ is larger for the former. In contrast to
the $L_{\rm X}$--$T_{\rm X}$ relation, the dispersion of the $c_\Delta$--$M_\Delta$
relation scatters the $L_{\rm X}$--$M_\Delta$ relation (dotted and dashed lines
in Figure~\ref{fig:LM}), which indicates that the $L_{\rm X}$--$M_\Delta$
relation is not an edge-on view of the fundamental plane.  
Moreover, as is the case of the $L_{\rm X}$--$T_{\rm X}$ relation, the
$L_{\rm X}$--$M_\Delta$ relation is written as $L_{\rm X}\propto
M_s^{(19+n)/12}\rho_s^{4/3}$ when only the fundamental plane relation is
used. Since $M_s$ is approximately represented by $M_s\propto
M_{500}^{1.1}$ (equation~(\ref{eq:MDMs})), the luminosity is given by
$L_{\rm X}\propto M_{500}^{1.56}\rho_s^{1.33}$ for $n=-2$, and $L_{\rm X}\propto
M_{500}^{1.51}\rho_s^{1.33}$ for $n=-2.5$. This means that the
$L_{\rm X}$--$M_\Delta$ relation is sensitive to $\rho_s$ and is scattered by
the variety.

We note that observationally determined directions of the fundamental
plane have some uncertainties caused by observational errors \citep[see
the contours in Figure~2 of][]{2018ApJ...857..118F}. On the other hand,
numerical simulations have shown that the plane is intrinsically thin
($\sim 0.03$~dex; \citealt{2018ApJ...857..118F}) and that the
thickness of the plane does not affect the plane normal. However, the
plane relation determined by the simulation results has a small
deviation from the relation we assumed in equation~(\ref{eq:TX})
\citep[see the marks in Figure~2 of][]{2018ApJ...857..118F}. They seem
to be associated with treatment of cool cores and presence or absence of
nongravitational effects. In order to estimate the influence of the
deviation, we construct fundamental plane relations $T_{\rm X}=T_{\rm X}(r_s, M_s)$
for each of the simulation sets (MUSIC, NF0, FB0, and FB1 in
\citealp{2018ApJ...857..118F}) instead of equation~(\ref{eq:TX}) and
derive the indices for the $L_{\rm X}$--$T_{\rm X}$ and the $L_{\rm X}$--$M_\Delta$
relations. We found that $1.4\lesssim \alpha \lesssim 1.9$ and
$1.1\lesssim \beta \lesssim 1.4$.  These uncertainties motivate us to
directly simulate the $L_{\rm X}$--$T_{\rm X}$ and $L_{\rm X}$--$M_\Delta$ relations in
the next section.

\begin{figure*}
\centering \plottwo{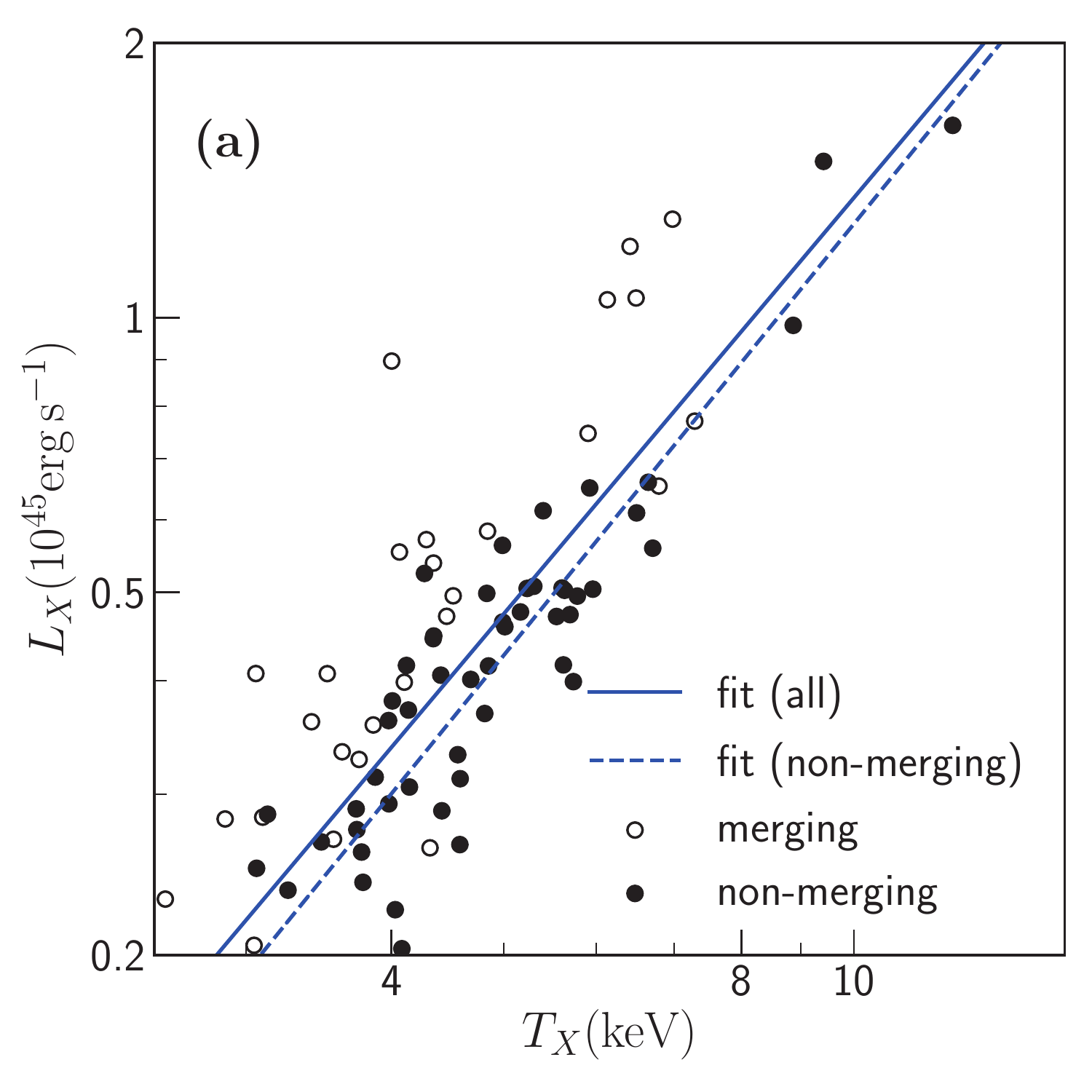}{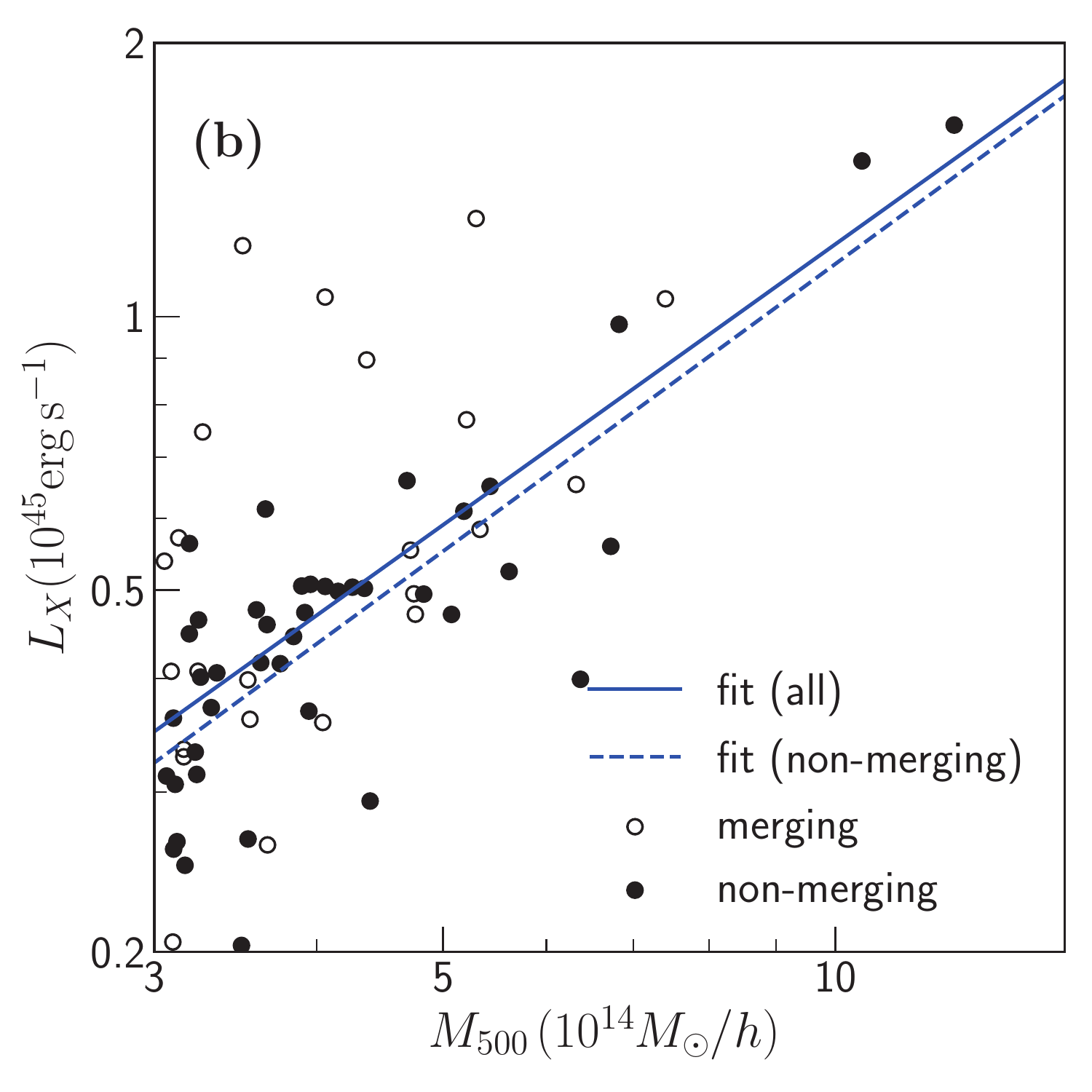}
\caption{Simulated (a) $L_{\rm X}$--$T_{\rm X}$, and (b) $L_{\rm X}$--$M_{500}$ relations
at $z=0$. The open circles are merging clusters and filled circles are
others (nonmerging clusters). The best fit for the whole simple and
that for the nonmerging clusters are shown by the solid and dashed
lines, respectively.\label{fig:LT_num}}
\end{figure*}

\section{Numerical simulations}
\label{sec:mum}

In order to confirm the predictions made in the previous section, we
analyze the results of the nonradiative version of the \textit{Omega500} hydrodynamical cosmological simulations \citep{nelson2014}. We do not include radiative cooling and feedback by
AGNs and supernovae when we calculate gas dynamics because the purpose
of this study is to find the baseline $L_{\rm X}$--$T_{\rm X}$ and $L_{\rm X}$--$M_\Delta$
relations.

The simulation \textit{Omega500} is run with Adaptive Tree Refinement,
an Eulerian code that uses adaptive refinement in space and time
\citep{kravtsov1999,kravtsov2002,rudd2008}.  The softening length is
$3.8\:h^{-1}$~kpc for both dark matter particles and gas cells in the
high-resolution regions.  Nonadaptive refinement in mass necessary for
resolving cores of the clusters is employed so that the highest mass
resolution for the dark matter particles is $m_{\rm DM}=1.09\times10^9\:
h^{-1}\: M_\odot$. The simulation box has a comoving box length of
$500\:h^{-1}$~Mpc. We select all of the 65 clusters at $z=0$ with
$M_{500}>3\times 10^{14}\: h^{-1}\: M_\odot$ regardless of dynamical
state. We compute the emission-weighted temperature including the
core. We kept the core because these simulations are nonradiative and
thus do not present cool-core features. The bolometric luminosity $L_{\rm X}$
and the temperature $T_{\rm X}$ are derived within a radius $r_{500}$. This
choice of the radius does not affect the results much if the radius is
large enough. This is because X-ray emissivity is proportional to
density squared and most of the X-ray emission comes from the central region
of clusters. The metal abundance is assumed to be $Z=0.3\: Z_\odot$ in
the calculation of the luminosity $L_{\rm X}$.

In Figure~\ref{fig:LT_num}, we show the results for the whole sample
(filled $+$ open circles). In Figure~\ref{fig:LT_num}(a), we fit the
data in log space with the function $L_{\rm X}\propto T_{\rm X}^\alpha$ using BCES
orthogonal regression (solid line; \citealp{1996ApJ...470..706A}). The
index for the fit is $\alpha=1.51\pm 0.09$ (all uncertainties are quoted at
the 1$\sigma$ confidence level unless otherwise mentioned), which is almost consistent
with the prediction for $n=-2$ in Section~\ref{sec:base}
(Table~\ref{tab:index}). However, some of the clusters in
Figure~\ref{fig:LT_num} are merging clusters, for which the ICM profiles
are often significantly deviated from the dark matter profiles. Thus, we
study the $L_{\rm X}$--$T_{\rm X}$ relation excluding merging clusters. We define
merging clusters as those that undergo a merger of mass ratio of at
least one-sixth in the past 2~Gyr, which is about a typical relaxation time
scale.  Figure~\ref{fig:LT_num}(a) shows that the dispersion of the
$L_{\rm X}$--$T_{\rm X}$ relation is reduced if we choose only nonmerging
clusters. The result of a fit for the nonmerging clusters is indicated
by the dashed line and the index is $\alpha=1.57\pm 0.08$. Changing the
merger mass ratio limit to one-fifth or one-seventh only shifts the slope within
$\pm 0.02$.  Again, the index is consistent with the prediction when
$n=-2$ (Table~\ref{tab:index}), and $\alpha=2$ is clearly rejected. For
the $L_{\rm X}$--$M_\Delta$ relation (Figure~\ref{fig:LT_num}(b)), the results
of fits show $\beta=1.02\pm 0.14$ for the whole sample and
$\beta=1.05\pm 0.12$ for the nonmerging clusters. Compared with the
$L_{\rm X}$--$T_{\rm X}$ relation (Figure~\ref{fig:LT_num}(a)), the dispersion is
larger, which may be related to the dispersion of the
$c_\Delta$--$M_\Delta$ relation (Figure~\ref{fig:LM}). The value of
$\beta$ is consistent with the predictions for $n=-2.5$ in
Table~\ref{tab:index} and $\beta=4/3\approx 1.33$ is clearly rejected.

%\begin{figure*}
%\centering \plottwo{f4a.eps}{f4b.eps} \caption{Examples of the
%realizations of the $f_{\rm gas}$--$T_{\rm X}$ relation. The error bars are
%observational uncertainties. (a) The assumed $\delta$ is 0.6 and the
%realized $\delta$ is $0.61\pm 0.07$. (b) The assumed $\delta$ is 0.45
%and the realized $\delta$ is $0.46\pm 0.04$. \label{fig:fgasT}}
%\end{figure*}

\section{Discussion}
\label{sec:dis}

We have shown that the suggested new relations for $L_{\rm X}$--$T_{\rm X}$
and $L_{\rm X}$--$M_\Delta$ once a mass profile is properly taken into account
are shallower than the ones predicted from a self-similar model.  With
the improved understanding of the luminosity-mass relation, we may explore
new methods of mass measurements through X-ray luminosity.  As shown in
Figure~\ref{fig:LM}, a typical scatter of 0.1~dex in the
$c_\Delta$--$M_\Delta$ relation can produce a scatter of approximately
0.15~dex in the $L_{\rm X}$--$M_\Delta$ relation. This will account for more
than 50\% of scatter in observed scatter
(e.g. \citealp{2007ApJ...668..772M,Rykoff2008,2008A&A...482..451Z}, but
see \citealp{2016A&A...585A.147A}). This suggests that that the scatter
caused by hierarchical structure formation can be comparable to
nongravitational effects, and proper modeling of this effect of halo
concentrations on the $L_{\rm X}$--$M_\Delta$ relation will help reduce the
scatter and potentially open the path for mass measurements through
X-ray luminosity.

If the shallow slopes of the baseline relations are observed, it will be
a proof of the hierarchical structure formation.  However, real clusters
are affected by the aforementioned additional nongravitational
effects. For example, the gas fraction of the central region of observed
clusters ($f_{\rm gas}$) is generally smaller than the baryon fraction
of the universe \citep[e.g.][]{2006ApJ...640..691V,2009ApJ...693.1142S}.
This is mostly explained as a result of the feedback from AGNs and
supernovae, although some of the baryon in clusters is consumed in star
formation. The feedback leads to an increase in $\alpha$ and $\beta$
from the baseline relations. Observations have shown that $\alpha\sim
2.6$--3.7 and $\beta\sim 1.6$--2.0
\citep[e.g.][]{1991MNRAS.252..414E,1996MNRAS.281..799E,1998ApJ...504...27M,1999MNRAS.305..631A,2002A&A...383..773I,2002ApJ...567..716R,2004A&A...417...13E,2008ApJ...680.1022H,2008A&A...482..451Z,2009A&A...498..361P,2011A&A...535A...4R}. Some
of the studies have shown that even if cluster cores are excised in the
analysis, the slopes (e.g. $\alpha= 2.8\pm 0.2$ and $\beta=1.63\pm
0.08$; \citealt{2007ApJ...668..772M}) are steeper than those of the
baseline relations ($\alpha\sim 1.7$ and $\beta\sim 1.1$--1.2). This
means that the feedback impacts on the ICM density beyond the cores. If
we assume that the indices of the observed relations are $\alpha=2.9$
and $\beta=1.8$ and those of the baseline relations are $\alpha=1.7$ and
$\beta=1.2$, the gas mass fraction at $r < r_s$ should have a temperature
and mass dependence of $f_{\rm gas}\propto T_{\rm X}^{(2.9-1.7)/2}=T_{\rm X}^{0.6}$
and $f_{\rm gas}\propto M_\Delta^{(1.8-1.2)/2}=M_\Delta^{0.3}$,
respectively. This is because the typical X-ray emissivity depends on
the gas fraction as $\epsilon\propto (\rho_s f_{\rm gas})^2\Lambda$. The
dependences are steeper than those expected in the simple self-similar
model ($f_{\rm gas}\propto T_{\rm X}^{(2.9-2)/2}=T_{\rm X}^{0.45}$ and $f_{\rm
gas}\propto M_\Delta^{(1.8-4/3)/2}\approx M_\Delta^{0.23}$ for
$\epsilon\propto (\rho_\Delta f_{\rm gas})^2\Lambda$; see
\citealp{2012A&A...539A.120B}).  It may be easier to confirm the $f_{\rm
gas}$--$T_{\rm X}$ relation than the $f_{\rm gas}$--$M_\Delta$ relation
because the former is steeper.  If these dependences are observationally
confirmed, they may indirectly prove the shallow baseline relations and
hierarchical structure formation.

The steeper slopes of the $f_{\rm gas}$--$T_{\rm X}$ and $f_{\rm
gas}$--$M_\Delta$ relations, when the new baseline relations are
adopted, can be explained as follows. The new baseline relations reflect
that low-temperature clusters have higher concentrations than
high-temperature clusters. Thus, the low-temperature clusters have
denser gas in the central region compared with those when they have the
same concentration as the high-temperature clusters.  Thus, stronger
feedback and smaller $f_{\rm gas}$ are required for the low-temperature
clusters in order to reproduce the observed steep $L_{\rm X}$--$T_{\rm X}$ and
$L_{\rm X}$--$M_\Delta$ relations. In the Appendix, we performed a mock analysis of
the $f_{\rm gas}$--$T_{\rm X}$ relation and showed that a sample of $\sim 20$
clusters is enough to discriminate $f_{\rm gas}\propto T_{\rm X}^{0.6}$ from
$f_{\rm gas}\propto T_{\rm X}^{0.45}$. This method to confirm the hierarchical
structure formation may be easier than observationally determining the
slope of the $c_\Delta$--$M_\Delta$ relation because the latter is
affected by a large scatter \citep[e.g.][]{2016MNRAS.461.3794O}.

We note that sample selection could be more important than the
number of clusters. For example, the observed $L_{\rm X}$--$T_{\rm X}$ and
$L_{\rm X}$--$M_\Delta$ relations could be affected by cluster mergers as well
as cool cores. In fact, \citet{2012MNRAS.421.1583M} derived a rather
small index is of $\alpha=1.90\pm 0.14$ for nonmerging clusters with
$T_{\rm X}\gtrsim 4$~keV when the emission from the cool cores are excised,
although a similar analysis done by \citet{2013ApJ...767..116M} showed
that $\alpha=2.26\pm 0.29$.\footnote{Previous studies that excise cool
cores often define the core as the region within $r=\chi\: r_\Delta$,
where $\chi$ is the constant (e.g. $\chi=0.15$ and $\Delta=500$;
\citealt{2012MNRAS.421.1583M,2013ApJ...767..116M}). If the
characteristic density of clusters is $r_s$ rather than $r_\Delta$, the
definition of the cool core should also be based on $r_s$. This
redefinition could change the slopes of the observed $L_{\rm X}$--$T_{\rm X}$ and
$L_{\rm X}$--$M_\Delta$ relations, although this is out of the scope of this
study.}  If the former is the case, it has already suggested the
shallower slope of the $L_{\rm X}$--$T_{\rm X}$ relation ($\alpha<2$), although
$\alpha=2$ cannot be rejected.  Since the gas fraction comes close to the
universal value for the highest-temperature clusters
\citep[e.g.][]{2006ApJ...640..691V}, the feedback appears to be less
effective for them and the shallower slope is more likely to be
realized. Thus, through the $L_{\rm X}$--$T_{\rm X}$ relation for
highest-temperature clusters, the hierarchical structure formation could
be proved without studying the $f_{\rm gas}$--$T_{\rm X}$ relation. In the near
future, \textit{eROSITA} would detect enough numbers of massive or
higher-temperature clusters and would enable us to confirm the shallow
$L_{\rm X}$--$T_{\rm X}$ and $L_{\rm X}$--$M_\Delta$ relations with a sufficient degree of
accuracy. Follow-up observations with {\it Chandra} and {\it XMM-Newton}
are also important.

The baseline relations we found should be taken into account especially
when the feedback effects are estimated based on the $L_{\rm X}$--$T_{\rm X}$ and/or
$L_{\rm X}$--$M_\Delta$ relations.  For example, our discovery indicates that
even if the observed index is proven to be $\alpha\sim 2$ or $\beta\sim
4/3$ after considering the effects of cluster mergers and cool cores, it
does not mean that those clusters are free from feedback because it is
still larger than that for the baseline relation ($\alpha\sim 1.7$ or
$\beta\sim 1.1$--$1.2$).

\section{Summary}
\label{sec:sum}

Using the mass dependence of halo concentrations and the fundamental
plane relation of galaxy clusters, we have shown that the index of the
baseline $L_{\rm X}$--$T_{\rm X}$ relation of clusters, that is, the one when
additional physics such as feedback, radiative cooling, and disturbance
by cluster mergers are ignored, should be $\alpha\sim 1.6$--1.8. The
value is smaller than $\alpha=2$, which was previously estimated based
on a simple self-similar model. For the baseline $L_{\rm X}$--$M_\Delta$
relation, we showed that the index should be $\beta\sim 1.1$--1.2, which
is also smaller than the prediction ($\beta=4/3$) of the self-similar
model.  These are because the halo concentration and the characteristic
density of clusters increase as the cluster mass decreases in the
hierarchical structure formation in a CDM universe. This mass dependence
was not considered when the conventional relations of $L_{\rm X}\propto T_{\rm X}^2$
and $L_{\rm X}\propto M_\Delta^{4/3}$ were derived. The new baseline relations
would be useful when the feedback effects are estimated based on scaling
relations. The baseline relations could be checked by the temperature or
mass dependence of gas mass fraction of clusters.  We also indicated
that the highest-temperature clusters may follow relations close to the new
baseline relations if the influences of cool cores and cluster mergers
are appropriately treated. This is because the feedback effects are
expected to be minimum for those clusters.  Near-future cluster surveys
(e.g. \textit{eROSITA}) may enable us to confirm the relations more
precisely. If they are actually confirmed, it would be a proof of the
hierarchical structure formation.

\acknowledgments

We thank the anonymous referee, whose comments improved the clarity of
this paper. We also thank Hans B\"ohringer and Daisuke Nagai for useful
comments and discussion. This work was supported by MEXT KAKENHI
No.~18K03647 (Y.F.).

\appendix

\section{Mock analysis of the gas fraction--temperature relation}
\label{sec:app}

For a quantitative discussion, we perform a simple mock analysis of the
$f_{\rm gas}$--$T_{\rm X}$ relation. We create a mock sample of $N$ clusters
in a temperature range of $T_{X,\rm min}\leq T_{\rm X} \leq T_{X,\rm max}$.
Since the number of clusters with larger temperatures is
smaller, we set the temperatures of the clusters following the equation
of $T_{\rm X}=(T_{X,\rm max} - T_{X,\rm min})x^2 + T_{X,\rm min}$, where $x$
is a random number between zero and one, and we assume that $T_{X,\rm
min}=4$~keV and $T_{X,\rm max}=12$~keV. As a result, the expected number
of clusters between 10 and 12~keV is about one-forth of that between 4 and
6~keV. The expected gas fraction at $r<\xi r_s$, where $\xi$ is an
appropriate constant, is assumed to be
\begin{equation}
\label{eq:fgas}
 f_{\rm gas} = f_{\rm gas,10}
\left(\frac{T_{\rm X}}{10\rm\: keV}\right)^\delta\:,
\end{equation}
where $f_{\rm gas,10}$ is the fraction for clusters with $T_{\rm X}=10$~keV,
and we assume $f_{\rm gas,10}=0.1$. The predicted indices for the new
and the conventional relations are $\delta=0.6$ and $\delta=0.45$,
respectively (Section~\ref{sec:dis}). For the conventional relation, we
implicitly assume that the halo concentration is independent of the
temperature and $r_s\propto r_\Delta$. Clusters distribute along the
relation~(\ref{eq:fgas}) with an intrinsic scatter. For the gas
fraction at $r<r_{500}$ or $r<r_{2500}$, observations have shown that
the intrinsic scatter is $\sim $10\% and observational uncertainties are
much smaller than the intrinsic scatter
\citep[e.g.][]{2006ApJ...640..691V}. As far as we know, there are no
previous observational studies on the gas fractions at $r<r_s$. The
determination of $r_s$ is rather difficult and the observational
uncertainties may be $\sim 30$\%
\citep[e.g.][]{2010A&A...524A..68E}. Fortunately, the gas fraction is
not sensitive to the radius, and even the 30\% uncertainties of the
radius cause $\lesssim 10$\% uncertainties for the gas fraction of
relaxed clusters (e.g. Figure~6 of \citealp{2013MNRAS.433.2790L}).
Thus, in addition to the 10\% intrinsic scatter, which is given by a
random gaussian, we introduce 10\% observational uncertainties for
$f_{\rm gas}$. We also assume 3\% observational uncertainties for $T_{\rm X}$.

For a given index $\delta$, we generate a total of $10^4$ realizations
of the sample. Using BCES$(f_{\rm gas}|T_{\rm X})$ regression
\citep{1996ApJ...470..706A}, we fit the data with
equation~(\ref{eq:fgas}) assuming that $\delta$ and $f_{\rm gas,10}$ are
free parameters, and derive the uncertainties of $\delta$. The results
are almost the same if we adopt a BCES orthogonal
regression. 
If we assume that $N=20$ and
$\delta=0.6$ (the new baseline prediction), the result of the
realizations is $\delta=0.61\pm 0.07$. This means
that $\delta=0.45$ (the conventional baseline prediction) can be
rejected with a sample of $N=20$ clusters. 
We note that $\delta=0.61\pm
0.10$ for $N=10$. 
On the other hand, if we assume that $\delta=0.45$ and $N=20$,
the result of the realizations is $\delta=0.45\pm 0.07$.

The above estimations suggest that the new baseline relation can be
confirmed once $f_{\rm gas}$ at $r<\xi r_s$ are derived for not too many
clusters. The radius $r_s$ can be determined by fitting an observed mass
profile with equation~(\ref{eq:MNFW}). Alternatively, if $M_\Delta$ at
two different $\Delta$ are obtained (say $\Delta=500$ and 2500), $r_s$
can also be derived from equations~(\ref{eq:MD}), (\ref{eq:cD}), and
(\ref{eq:MNFW}).




\begin{thebibliography}{}

\bibitem[Adami et al.(1998)]{1998A&A...331..493A} Adami, C., Mazure, A.,
		Biviano, A., Katgert, P., \& Rhee, G.\ 1998, \aap, 331,
		493

\bibitem[Akritas \& Bershady(1996)]{1996ApJ...470..706A} Akritas, M.~G.,
		\& Bershady, M.~A.\ 1996, \apj, 470, 706

\bibitem[Andreon et al.(2016)]{2016A&A...585A.147A} Andreon, S., Serra,
		A.~L., Moretti, A., \& Trinchieri, G.\ 2016, \aap, 585,
		A147

\bibitem[Araya-Melo et al.(2009)]{2009MNRAS.400.1317A} Araya-Melo,
		P.~A., van de Weygaert, R., \& Jones, B.~J.~T.\ 2009,
		\mnras, 400, 1317

\bibitem[Arnaud \& Evrard(1999)]{1999MNRAS.305..631A} Arnaud, M., \&
		Evrard, A.~E.\ 1999, \mnras, 305, 631

\bibitem[Bertschinger(1985)]{1985ApJS...58...39B} Bertschinger, E.\
		1985, \apjs, 58, 39

\bibitem[Bhattacharya et al.(2013)]{2013ApJ...766...32B} Bhattacharya,
		S., Habib, S., Heitmann, K., \& Vikhlinin, A.\ 2013,
		\apj, 766, 32

\bibitem[Biviano \& Salucci(2006)]{2006A&A...452...75B} Biviano, A., \&
		Salucci, P.\ 2006, \aap, 452, 75

\bibitem[B{\"o}hringer et al.(2012)]{2012A&A...539A.120B} B{\"o}hringer,
		H., Dolag, K., \& Chon, G.\ 2012, \aap, 539, A120

\bibitem[Borgani et al.(2004)]{2004MNRAS.348.1078B} Borgani, S.,
		Murante, G., Springel, V., et al.\ 2004, \mnras, 348,
		1078

\bibitem[Bryan \& Norman(1998)]{1998ApJ...495...80B} Bryan, G.~L., \&
		Norman, M.~L.\ 1998, \apj, 495, 80

\bibitem[Child et al.(2018)]{2018ApJ...859...55C} Child, H.~L., Habib,
		S., Heitmann, K., et al.\ 2018, \apj, 859, 55

\bibitem[Correa et al.(2015)]{2015MNRAS.452.1217C} Correa, C.~A.,
		Wyithe, J.~S.~B., Schaye, J., \& Duffy, A.~R.\ 2015,
		\mnras, 452, 1217

\bibitem[Diemer \& Kravtsov(2015)]{2015ApJ...799..108D} Diemer, B., \&
		Kravtsov, A.~V.\ 2015, \apj, 799, 108

\bibitem[Dvorkin \& Rephaeli(2015)]{2015MNRAS.450..896D} Dvorkin, I., \&
		Rephaeli, Y.\ 2015, \mnras, 450, 896

\bibitem[Duffy et al.(2008)]{2008MNRAS.390L..64D} Duffy, A.~R., Schaye,
		J., Kay, S.~T., \& Dalla Vecchia, C.\ 2008, \mnras, 390,
		L64

\bibitem[Ebeling et al.(1996)]{1996MNRAS.281..799E} Ebeling, H., Voges,
		W., B\"ohringer, H., et al.\ 1996, \mnras, 281, 799

\bibitem[Edge \& Stewart(1991)]{1991MNRAS.252..414E} Edge, A.~C., \&
		Stewart, G.~C.\ 1991, \mnras, 252, 414

\bibitem[Eisenstein \& Hu(1998)]{1998ApJ...496..605E} Eisenstein, D.~J.,
		\& Hu, W.\ 1998, \apj, 496, 605

\bibitem[Enoki et al.(2001)]{2001ApJ...556...77E} Enoki, M., Takahara, F., \& Fujita, Y.\ 2001, \apj, 556, 77 

\bibitem[Ettori(2013)]{2013MNRAS.435.1265E} Ettori, S.\ 2013, \mnras, 435, 1265 

\bibitem[Ettori(2015)]{2015MNRAS.446.2629E} Ettori, S.\ 2015, \mnras, 446, 2629 


\bibitem[Ettori et al.(2010)]{2010A&A...524A..68E} Ettori, S.,
		Gastaldello, F., Leccardi, A., et al.\ 2010, \aap, 524,
		A68

\bibitem[Ettori et al.(2004)]{2004A&A...417...13E} Ettori, S., Tozzi,
		P., Borgani, S., \& Rosati, P.\ 2004, \aap, 417, 13

\bibitem[Fujita et al.(2019)]{2019arXiv190100008F} Fujita, Y., Donahue,
		M., Ettori, S., et al.\ 2019, Galaxies, 7, 8
		(arXiv:1901.00008).

\bibitem[Fujita \& Takahara(1999)]{1999ApJ...519L..51F} Fujita, Y., \&
Takahara, F.\ 1999, \apjl, 519, L51

\bibitem[Fujita et al.(2018a)]{2018ApJ...857..118F} Fujita, Y., Umetsu,
K., Rasia, E., et al.\ 2018a, \apj, 857, 118

\bibitem[Fujita et al.(2018b)]{2018ApJ...863...37F} Fujita, Y., Umetsu,
		K., Ettori, S., et al.\ 2018b, \apj, 863, 37

\bibitem[Fujita \& Ohira(2013)]{2013MNRAS.428..599F} Fujita, Y., \&
		Ohira, Y.\ 2013, \mnras, 428, 599

\bibitem[Gonzalez et al.(2013)]{2013ApJ...778...14G} Gonzalez, A.~H.,
		Sivanandam, S., Zabludoff, A.~I., et al.\ 2013, \apj,
		778, 14.

\bibitem[Hicks et al.(2008)]{2008ApJ...680.1022H} Hicks, A.~K.,
		Ellingson, E., Bautz, M., et al.\ 2008, \apj, 680, 1022

\bibitem[Ikebe et al.(2002)]{2002A&A...383..773I} Ikebe, Y., Reiprich,
		T.~H., B{\"o}hringer, H., Tanaka, Y., \& Kitayama, T.\
		2002, \aap, 383, 773

\bibitem[Kaiser(1986)]{1986MNRAS.222..323K} Kaiser, N.\ 1986, \mnras,
		222, 323

\bibitem[Kravtsov(1999)]{kravtsov1999} Kravtsov A. V.\ 1999, PhD thesis, New Mexico State University

\bibitem[Kravtsov et al. (2002)]{kravtsov2002} Kravtsov A. V., Klypin
		A., Hoffman Y.\ 2002, \apj, 571, 563

\bibitem[Landry et al.(2013)]{2013MNRAS.433.2790L} Landry, D.,
		Bonamente, M., Giles, P., et al.\ 2013, \mnras, 433,
		2790.

\bibitem[Lanzoni et al.(2004)]{2004ApJ...600..640L} Lanzoni, B., Ciotti,
		L., Cappi, A., Tormen, G., \& Zamorani, G.\ 2004, \apj,
		600, 640

\bibitem[Ludlow et al.(2013)]{2013MNRAS.432.1103L} Ludlow, A.~D.,
		Navarro, J.~F., Boylan-Kolchin, M., et al.\ 2013,
		\mnras, 432, 1103

\bibitem[Mahdavi et al.(2013)]{2013ApJ...767..116M} Mahdavi, A.,
		Hoekstra, H., Babul, A., et al.\ 2013, \apj, 767, 116

\bibitem[Markevitch(1998)]{1998ApJ...504...27M} Markevitch, M.\ 1998,
		\apj, 504, 27

\bibitem[Maughan(2007)]{2007ApJ...668..772M} Maughan, B.~J.\ 2007, \apj,
		668, 772

\bibitem[Maughan(2014)]{2014MNRAS.437.1171M} Maughan, B.~J.\ 2014, \mnras, 437, 1171 

\bibitem[Maughan et al.(2012)]{2012MNRAS.421.1583M} Maughan, B.~J.,
		Giles, P.~A., Randall, S.~W., Jones, C., \& Forman,
		W.~R.\ 2012, \mnras, 421, 1583

\bibitem[Meneghetti et al.(2014)]{2014ApJ...797...34M} Meneghetti, M.,
		Rasia, E., Vega, J., et al.\ 2014, \apj, 797, 34

\bibitem[Navarro et al.(1997)]{1997ApJ...490..493N} Navarro, J.~F.,
		Frenk, C.~S., \& White, S.~D.~M.\ 1997, \apj, 490, 493
		
\bibitem[Nelson et al.(2014)]{nelson2014} Nelson K., Lau E. T., Nagai
		D., Rudd D. H., Yu L.\ 2014, \apj, 782, 107

\bibitem[Okabe \& Smith(2016)]{2016MNRAS.461.3794O} Okabe, N., \& Smith,
		G.~P.\ 2016, \mnras, 461, 3794

\bibitem[Ota et al.(2006)]{2006ApJ...640..673O} Ota, N., Kitayama, T.,
		Masai, K., \& Mitsuda, K.\ 2006, \apj, 640, 673

\bibitem[Peebles(1980)]{1980lssu.book.....P} Peebles, P.~J.~E.\ 1980,
		The large-scale structure of the universe (Princeton, NJ: Princeton Univ. Press)

\bibitem[Pratt et al.(2009)]{2009A&A...498..361P} Pratt, G.~W., Croston,
		J.~H., Arnaud, M., \& B{\"o}hringer, H.\ 2009, \aap,
		498, 361
		
\bibitem[Puchwein et al.(2008)]{2008ApJ...687L..53P} Puchwein, E.,
		Sijacki, D., \& Springel, V.\ 2008, \apjl, 687, L53

\bibitem[Reichert et al.(2011)]{2011A&A...535A...4R} Reichert, A.,
		B{\"o}hringer, H., Fassbender, R., \& M{\"u}hlegger, M.\
		2011, \aap, 535, A4

\bibitem[Reiprich \& B{\"o}hringer(2002)]{2002ApJ...567..716R} Reiprich,
		T.~H., \& B{\"o}hringer, H.\ 2002, \apj, 567, 716

\bibitem[Rudd et al.(2008)]{rudd2008} Rudd D. H., Zentner A. R., Kravtsov A. V.\ 2008, \apj, 672, 19

\bibitem[Rykoff et al.(2008)]{Rykoff2008} Rykoff, E. S. and Evrard, A. E. and McKay, T. A.\ 2008, \mnras, 387, 28

\bibitem[Schaeffer et al.(1993)]{1993MNRAS.263L..21S} Schaeffer, R.,
		Maurogordato, S., Cappi, A., \& Bernardeau, F.\ 1993,
		\mnras, 263, L21

\bibitem[Sutherland \& Dopita(1993)]{1993ApJS...88..253S} Sutherland,
		R.~S., \& Dopita, M.~A.\ 1993, \apjs, 88, 253

\bibitem[Sun et al.(2009)]{2009ApJ...693.1142S} Sun, M., Voit, G.~M.,
		Donahue, M., et al.\ 2009, \apj, 693, 1142

\bibitem[Verde et al.(2002)]{2002ApJ...581....5V} Verde, L., Haiman, Z.,
		\& Spergel, D.~N.\ 2002, \apj, 581, 5

\bibitem[Vikhlinin et al.(2006)]{2006ApJ...640..691V} Vikhlinin, A.,
		Kravtsov, A., Forman, W., et al.\ 2006, \apj, 640, 691

\bibitem[Voit et al.(2002)]{2002ApJ...576..601V} Voit, G.~M., Bryan,
		G.~L., Balogh, M.~L., \& Bower, R.~G.\ 2002, \apj, 576,
		601

\bibitem[Wechsler et al.(2002)]{2002ApJ...568...52W} Wechsler, R.~H.,
		Bullock, J.~S., Primack, J.~R., Kravtsov, A.~V., \&
		Dekel, A.\ 2002, \apj, 568, 52

\bibitem[Zhang et al.(2008)]{2008A&A...482..451Z} Zhang, Y.-Y.,
		Finoguenov, A., B{\"o}hringer, H., et al.\ 2008, \aap,
		482, 451

\end{thebibliography}
\end{document}